# Effect of gallium doping on structural and transport properties of the topological insulator $Bi_2Se_3$ grown by molecular beam epitaxy

Daniel Brito[1,2], Ana Pérez-Rodriguez[3], Ishwor Khatri[1], Carlos José Tavares[4], Mario Amado[3], Eduardo Castro[5], Enrique Diez[3], Sascha Sadewasser[1,6] and Marcel S. Claro[1,6*]

1. International Iberian Nanotechnology Laboratory, Av. Mestre Jose Veiga, Braga 4715-330, Portugal
2. Departamento de Física e Astronomia, Faculdade de Ciências, Universidade do Porto, rua do Campo Alegre s/n, 4169-007 Porto, Portugal
3. Nanotechnology Group, Department of Fundamental Physics, University of Salamanca, 37008 Salamanca, Spain
4. Physics Centre of Minho and Porto Universities (CF-UM-PT), University of Minho, Azurém, 4804-533 Guimarães, Portugal
5. Centro de Física das Universidades do Minho e Porto (CF-UM-PT), Departamento de Físisca e Astronomia, Faculdade de Ciências, Universidade do Porto, 4169-007 Porto, Portugal
6. QuantaLab, 4715-330 Braga, Portugal

*corresponding author: marcel.claro@inl.int

**Abstract**

Topological insulators possess a non-conductive bulk and present surface states, henceforth, they are electrically conductive along their boundaries. Bismuth selenide ($Bi_2Se_3$) is one of the most promising topological insulators. However, a major drawback is its n-type nature arising from its natural doping, which makes the transport in the bulk dominant. This effect can be overcome by shifting the chemical potential into the bandgap, turning the transport of the surface states to be more pronounced than the bulk counterpart. In this work, $Bi_2Se_3$ was grown by molecular beam epitaxy and doped with 0.8, 2, 7, and 14 at. % of Ga, with the aim of shifting the chemical potential into the bandgap. The structural, morphological, and electronic properties of the Ga doped $Bi_2Se_3$ are studied. Raman and X-ray diffraction measurements confirmed the incorporation of the dopants into the crystal structure. Transport and magnetoresistance measurements in the temperature range of 1.5 to 300 K show that Ga-doped $Bi_2Se_3$ is n-type with a bulk charge carrier concentration of $10^{19}$ $cm^{-3}$. Remarkably, magnetotransport of the weak antilocalization effect (WAL) measurements confirm the existence of surface states up to a doping percentage of 2 at. % of Ga and coherence length values between 50-800 nm, which envisages the possibility of topological superconductivity in this material.

**Author Keywords.** *Bi*$_2$*S*e$_3$, Ga-doping, surface states, topological insulator, weak antilocalization effect.

## 1. Introduction

Topological insulators (TI) are a type of quantum materials that feature robust surface states protected by time-reversal symmetry (TRS)[1]. Their robustness implies that there is no backscattering allowed, and therefore there is no energy loss, making these materials promising candidates for low dissipation transport applications[1].

$Bi_2Se_3$ is a 3D TI characterized by a robust single Dirac-cone of topological surface states, which crosses the bandgap of 0.3 eV [2], making them one of the most suitable TIs for studying the topologically protected states at room temperature, since the thermal energy will be lower than its bandgap. Unfortunately, $Bi_2Se_3$ typically grows n-type degenerated (n≈$10^{19}$ $cm^{-3}$) due to defects in the crystal, such as Se vacancies[1,3], which moves the Fermi level into the conduction band. Moreover, in order to exploit the surface states properties, it is of fundamental importance to reduce the bulk carrier concentration, with the purpose to avoid bulk contributions to carrier transport. One possible approach is p-type doping of the material to shift the chemical potential into the bandgap, closer to the Dirac-cone. Many studies have been focused on tuning the chemical potential of $Bi_2Se_3$ by doping with Co[2], Ag[3,4], In[5,6], and Cu[7,8]. The study of different dopants is fundamental, looking for the possibility of tuning the chemical potential in the bandgap or promoting the discovery of new physical properties.

In this work, we focus on the effect of doping $Bi_2Se_3$ with Ga. An illustration of the crystal structure of $Bi_2Se_3$ is presented in Fig. 1[9]. The rhombohedral crystal structure is known as a stack of quintuple layers, which consist of five atomic layers ordered as Se-Bi-Se-Bi-Se. These quintuple layers are bonded by Van der Waals forces[10]. Ga can be inserted into the crystal structure in three possible ways: by replacing Bi atoms in quintuple layers (Fig. 1b); in interstitial positions in the quintuple layers sites (Fig. 1c); or intercalated in the Van der Waals gap (Fig. 1d). In a recent theoretical study, Ga doping of $Bi_2Se_3$ was shown to lead to in-gap resonant level peaks in the density of states, which suggests a possible impurity for p-type doping[11]. However, the calculations were done in the coherent potential approximation, which does not allow to distinguish between the three crystal structure positions for Ga. In addition Phutela et al. studied new TIs, demonstrating that $GaBiSe_2$ is a strong topological insulator[12]. Thus, in order to fully explore its potential, experimental information about Ga-doped $Bi_2Se_3$ is scarce and further research is needed.

Motivated to fill the knowledge gap, this work provides a thorough study of the influence of Ga as a dopant for $Bi_2Se_3$. In a systematic study, we create a correlation between the structural and electrical properties. The material was deposited by molecular beam epitaxy (MBE), co-evaporating the elements Bi, Se, and Ga on c-oriented sapphire substrates. Structural properties, transport properties, magnetoresistance, and low-temperature magnetoresistance were investigated in order to determine the influence of Ga doping with nominal doping levels up to 14 at.%. This study adds insight on the crystal structure of the material and the possible arrangement of Ga atoms.

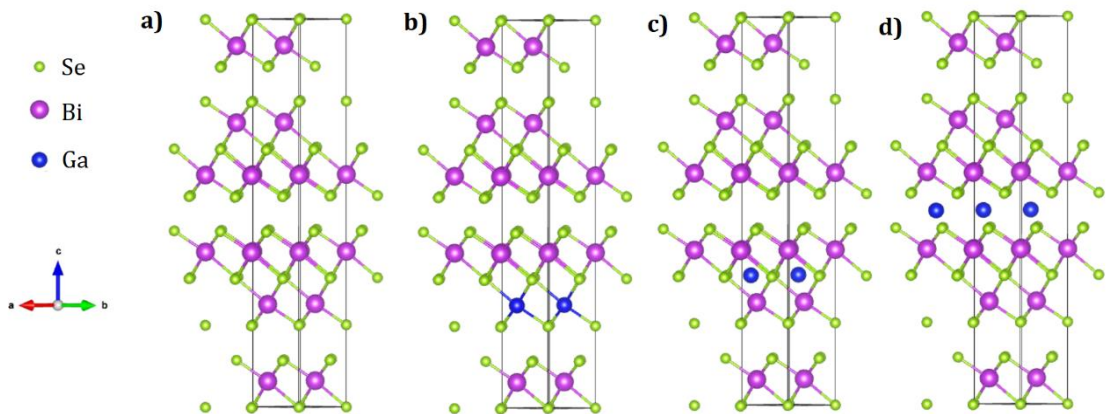

**Figure 1.** Crystal structure of a) $Bi_2Se_3$, and possible insertion configurations of Ga in the crystal structure as b) Bi substitution, c) interstitial into the quintuple layer or d) intercalation into the Van der Waals gap[9].

## 2. Materials and Characterization Techniques

### a. Growth Procedure

Ga$_x$$Bi_2$Se$_3$ films were grown onto c-axis oriented sapphire substrates by molecular beam epitaxy (MBE, Omicron Nanotechnology EVO-50). The substrates were heated up to 950 °C prior to the growth to remove oxides and any other undesired species. In-situ reflection high energy electron diffraction (RHEED) was used to observe the crystal surface quality. Bi and Se were evaporated at 535 °C and 285 °C, respectively. Ga was evaporated at temperatures varying from 680 ºC to 765 ºC. Furthermore, a cracker stage at 900 ºC was used in the Se cell to break large molecules of Se into smaller ones. High purity materials of 99.999% were used (AlfaAesar). The ratio of Bi to Se was kept at 1:4, which was reported to be the preferential ratio[13], leading to a growth rate of ≈1.4 nm/min of $Bi_2$Se$_3$. The substrate temperature during growth was fixed at 310 ºC. Since the temperature of the Se cell is lower than the substrate temperature, the growth rate is relying only on the flux of Bi[14]. Therefore, the doping was controlled by varying Ga:Bi flux ratio (8:1000 (0.8 at.%), 1:50 (2 at.%), 7:100 (7 at.%), and 7:50 (14 at.%)), upon increasing the temperature of the Ga cell. Before the growth, a surface treatment was performed by growing the first few layers at a low temperature (140 °C) (further details of the surface treatment can be found in *I. Levy 2018* [15]). Then, the substrate temperature was increased to 310 °C for the growth of $Bi_2$Se$_3$ at a pressure of 10$^{-7}$ mbar. The growth time was 40 min for all samples, aiming for 60 nm thickness of $Bi_2$Se$_3$ (samples of Ga:Bi = 8:1000 (0.8 at.%) have a thickness of 40 nm due to a change of growth rate over time).

### b. XPS

The composition of the samples was determined by X-ray photoelectron spectroscopy (XPS), using an ESCALAB250Xi ultra-high vacuum (UHV) system (Thermo Fisher Scientific). All samples were treated with a plasma etching for 2 min with Ar$^+$ ions to remove any surface oxides. XPS spectra were used to quantify the atomic composition of the samples, for which the areas of the peaks of all elements were fitted and integrated by the data processing software Avantage (Thermo Fisher Scientific), using sensitivity factors provided by Avantage library.



### c. X-ray diffraction and Raman

The crystal structure analysis was performed by X-ray diffraction (XRD, X'Pert Panalytical). The film thickness was obtained by X-ray reflection analysis. The analysis of the vibrational modes was done employing a confocal Raman microscopy system (WITec alpha300 R) equipped with an optical microscope. The spectroscopy was performed with a laser wavelength of 532 nm and a power of approximately 0.8 mW to prevent sample damage and a 1800 grating and resolution of 0.1 cm$^{-1}$.

### d. Magnetoresistance

Transport properties such as resistivity, mobility, and electron density were measured on a Hall system (TESLATRON USAL) within a temperature range of 1.5 - 300 K in a four-point Van der Pauw geometry with indium contacts, under a magnetic field up to 5T. The low-temperature measurements were made by a standard lock-in technique with frequencies of 10 - 20 Hz and a driving current of 0.5 µA to guarantee both low heating and sufficient signal.

## 3. Results and Discussion

In-situ RHEED results of the samples are shown in Figure 2. The good $Bi_2Se_3$ quality is confirmed by sharp streaky patterns and a six-fold symmetry with sample rotation (Fig. 2 a), corresponding to a flat surface with small domains. RHEED patterns are consistent up to Ga concentrations of 2 at.% (Fig. 2 b). The sample with 0.8 at.% Ga (not shown) does not show any variation in the $Bi_2Se_3$ RHEED pattern, compared with the undoped $Bi_2Se_3$. For 7 at.% Ga, the RHEED exhibits some spots (Fig. 2 c), which are attributed to an increase in roughness produced by the Ga incorporation in the film. Despite this, the $Bi_2Se_3$ pattern is clearly identified, implying that the crystal quality is maintained. As the concentration of Ga is increased, concentric rings start to appear (Fig. 2 d), which are related to an increasing polycrystallinity of the film. Additionally, the RHEED pattern becomes weaker, which is attributed to the poor crystallinity of the film, as also confirmed by the XRD analysis (see below Fig. 4).



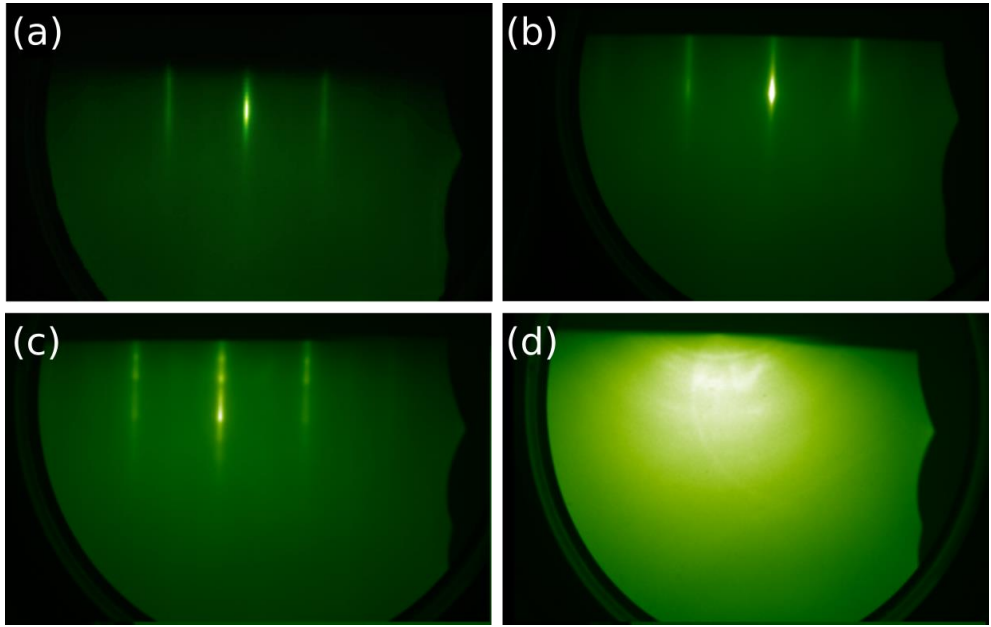

**Figure 2.** RHEED images of a) undoped $Bi_2Se_3$ and doped $Bi_2Se_3$ with b) 2 at.%, c) 7 at.%, and d) 14 at.% Ga.

XPS was employed to study the doping level. Fig. 3 shows the high-resolution XPS spectra of the $Bi_2Se_3$ films with increasing Ga doping. The appearance of the peaks corresponding to the binding energies (BE) of Ga confirms the doping of $Bi_2Se_3$ (Fig. 3a). It is observed that the peak represented by *Bi 4f* (Fig. 3b) does not show the presence of oxides and impurities, such as $Bi_2O_3$[16]. The BE of the *Bi $4f_{7/2}$* and *$4f_{5/2}$* doublet peaks are localized at 158.10 eV and 163.40 eV, respectively. The *Se 3d* doublet separation is not discerned due to the close positions of the *$3d_{5/2}$* and *$3d_{3/2}$* peaks at ~54 eV (Fig. 3c). Likewise, the BE of *Ga 3d* doublet peaks are superposed at ~20 eV.

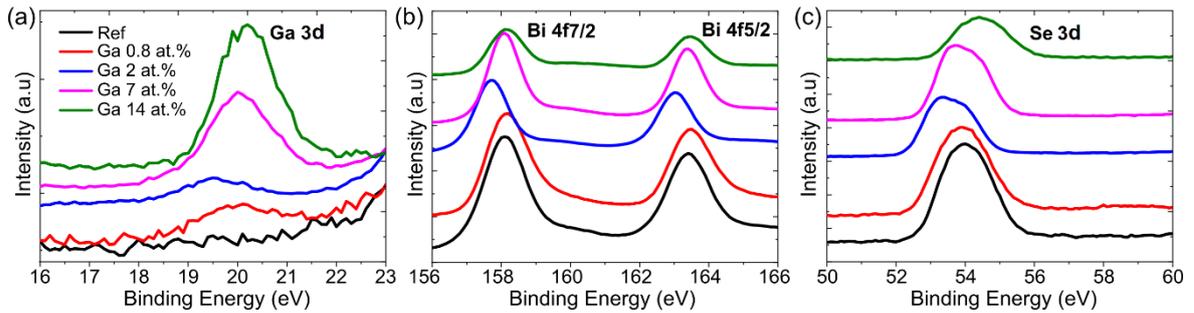

**Figure 3.** High-resolution XPS spectra for a) *Ga 3d,* b) *Bi 4f*, c) *Se 3d* core lines.

From the XPS spectra the atomic composition of the samples was quantified (Table I). The data shows a good agreement between the nominal and measured values of the Ga doping. Despite the small differences, the nominal doping concentrations in the following analysis were used.



**Table I.** Atomic concentrations of Bi, Se, and Ga derived from XPS analysis for all samples.

| Elements | Chemical composition (at.%) | | | | |
|---|---|---|---|---|---|
| | Undoped $Bi_2Se_3$ | 0.8 at.% Ga | 2 at.% Ga | 7 at.% Ga | 14 at.% Ga |
| Bi | 33.4 | 35.3 | 31.8 | 29.6 | 21.1 |
| Se | 66.6 | 63.9 | 66.8 | 64.5 | 65.5 |
| Ga | -- | 0.8 | 1.4 | 5.9 | 13.4 |

The crystal structure of the material was studied by XRD. The diffractograms of all samples are shown in Fig. 4 a), with the nominal Ga-doping percentage increasing from the bottom to the top. All samples present a single-crystal pattern of $Bi_2Se_3$, corresponding to a rhombohedral crystal structure (R-3m), which is consistent with the literature[15]. The pattern shows a {00L} diffraction family, indicating a preferential epitaxial growth along the c-axis of the substrate without impurities, such as GaSe[17]. These peaks shift towards lower Bragg angles, as the dopant percentage is increased, indicating a lattice parameter variation with the doping concentration. Fig. 4b) shows the high-resolution XRD data for the (0015) diffraction peak. As the doping level is increased the peak intensity is reduced, suggesting a deterioration of the crystal quality. For 2 and 7 at.% Ga doping, a double peak appears, indicating a possible formation of a secondary phase. This peak split is shifted toward lower Bragg angles, corresponding to a higher c-lattice parameter. For 14 at.% Ga, again a single peak is observed, shifted to a lower Bragg angle compared with undoped $Bi_2Se_3$.

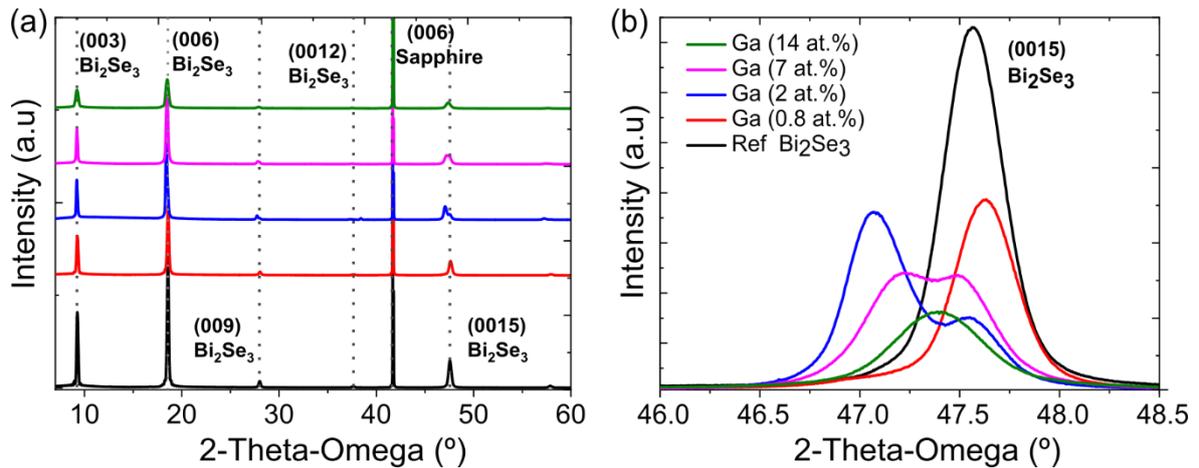

**Figure 4.** a) XRD diffractogram for the $Bi_2Se_3$ doped with Ga and b) High resolution XRD diffractogram for the peak (0015).



The crystal lattice constants were calculated according to:

$$\frac{1}{d_{hkl}^2} = \frac{4(h^2+hk+k^2)}{3a^2} + \frac{l^2}{c^2} , \qquad (1)$$

for hexagonal crystal structures and are presented as a function of dopant concentration in Fig. 5. The value of *a=b* was obtained from the analysis of the *Bi*$_2$*Se*$_3$ (015) diffraction peak (S.I. 2). Additionally, a fitting was performed by the Fityk software to obtain a higher precision (dots in Fig. 5).

The lattice parameters of undoped *Bi*$_2$*Se*$_3$ agree with typical literature values (*a*=4.139 Å and *c*=28.66 Å)[18–20]. As Ga$^{+3}$ has a smaller ionic radius (0.62 Å) than Bi$^{3+}$ (1.03 Å), the incorporation of Ga by substituting of Bi atoms should lead to a decrease in the crystal lattice[21]. However, the lattice constants *c* and *a* increases gradually with increasing Ga content, indicating that the volume of the unit cell expands, which is consistent with the intercalation of Ga atoms in the Van der Waals gap[22]. These results are in agreement with the simulated calculations by density perturbation theory (DFT) (Table S.1). The lattice parameter c(Å), in terms of percentage, shows an increase of 0.7% for the sample with 2 at.% Ga doping, which is similar to the simulations of Ga atoms vdW intercalated, where a 1.2% increase was obtained. The simulations for the lattice constant a(Å) also reveal an increase upon Ga intercalation, similar to the experimental results. Similarly, studies on Cu/Ag/Co-doped *Bi*$_2$*Se*$_3$ show an increase in the c-lattice parameter due to the intercalation of dopant atoms in Van der Waals gap[2,4,7,23,24]. In contrast, other studies have reported that In incorporation results in a decrease in the value of c, apparently indicating the substitution of Bi atoms [19,25]. Presumably, these results restrain the possibility of a substitution of Bi atoms with Ga. Even though, for 14 at.% Ga doping, the lattice parameters a and c show a decrease to 4.137Å and 28.77Å, respectively, which are attributed to a mixture of intercalated and substitutional Ga atoms. The high flux of Ga atoms would lead to a replacement of Bi atoms with Ga resulting in a reduction of the lattice parameter a [7].



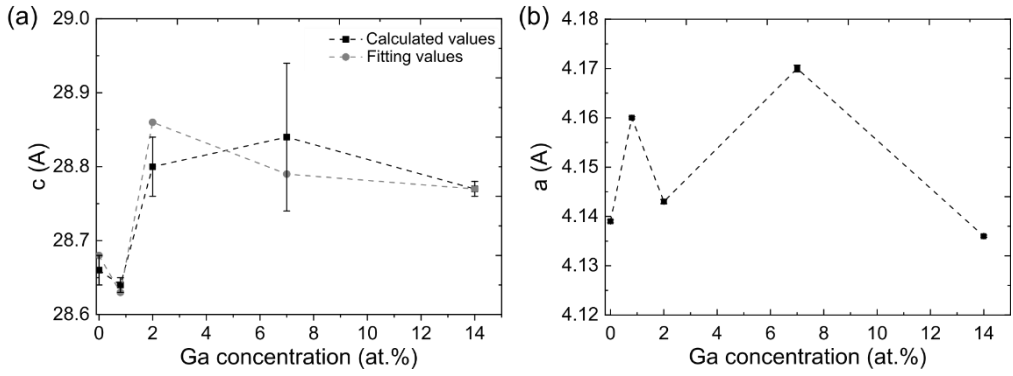

**Figure 5.** Crystal lattice parameter a) *c* and b) *a* as a function of the Ga doping.

The dopant influence on the vibrational modes of $Bi_2Se_3$ was analyzed by Raman spectroscopy at room temperature (Fig. 6 a). Three characteristic $Bi_2Se_3$ Raman peaks are observed, belonging to $A_{1g}^1$, $E_g^2$, and $A_{1g}^2$ vibrational modes, at 71 cm$^{-1}$, 130 cm$^{-1}$, 174 cm$^{-1}$, respectively [2,26]. A gradual redshift in the vibrational Raman peaks is observed as the Ga concentration increases in the material, which we can clearly attribute to the Ga incorporation into the crystal structure, as no other material phases or contamination sources were detected.

Fig 6 b) shows the peak position for all the vibrational modes, as a function of the nominal dopant element concentration. Recent studies of Cu doping of $Bi_2Se_3$ also showed a similar redshift with an increasing Cu concentration[7], concluding that a shift to lower frequencies is related to the intercalation of atoms in the crystal structure. Thus, it seems reasonable to assume that Ga atoms are intercalated within the Van der Walls gap in the crystal structure, corroborating the conclusions from the XRD analysis. Oppositely, a shift to higher frequencies is due to the replacement of Bi atoms with lighter atoms, which leads to stronger bonding forces [21,23]. From these outcomes, it is possible to assume that Ga atoms are not replacing the Bi atoms effectively since Ga is a lighter element than Bi for higher concentrations of Ga than 0.8 at.%. For the sample doped with 0.8 at.% of Ga, a blueshift is observed, indicating the replacement of Bi atoms for a lower flux of Ga.



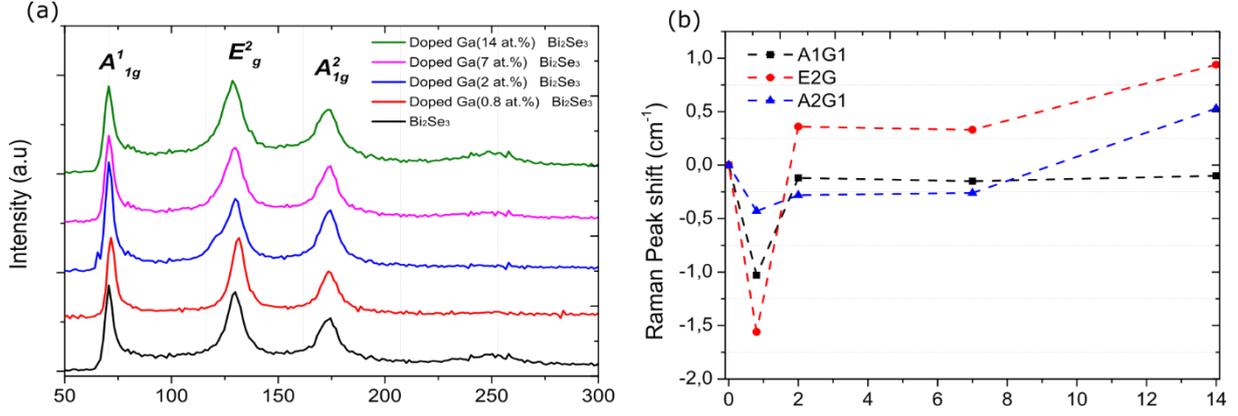

**Figure 6.** a) Raman spectra of undoped $Bi_2Se_3$ and $Bi_2Se_3$ doped with varying Ga concentration. b) Raman peak position evolution as a function of the nominal Ga-concentration in $Bi_2Se_3$.

In order to characterize the level of defects and their impact on the transport properties, we performed Hall effect measurements in the samples. Fig. 7 shows the resistivity, bulk and sheet concentrations, and electron mobility as a function of temperature. To understand the effect of doping on the surface transport properties, compared to undoped $Bi_2Se_3$, we measured samples with Ga doping of 2 at.% and 7 at.% and the undoped $Bi_2Se_3$ (room temperature results for 0.8 at.% and 14 at.% are shown in Fig. S.I 3). Undoped $Bi_2Se_3$ presents a sheet concentration of $\sim 7 \times 10^{13}$ cm$^{-2}$ and mobility of 423.5 cm$^2$ cm$^2$V$^{-1}$s$^{-1}$, in good agreement with literature values for MBE-grown $Bi_2Se_3$[1,10,27]. Ga doping increases the bulk concentration almost 6-fold, from $1.1 \times 10^{19}$ cm$^{-3}$ to $\sim 6 \times 10^{19}$ cm$^{-3}$, while the mobility decreases to 119.2 cm$^2$ cm$^2$V$^{-1}$s$^{-1}$. The observed increase of the electron concentration suggests that Ga (replacing Bi atoms) is not acting as an acceptor impurity. However, for the doping level of 14 at.% the bulk concentration increases only 3-fold, which is in agreement with the assumed mixture of intercalated and substitutional Ga atoms. Furthermore, when an increased number of defects is introduced into the Van der Waals gap (intercalated atoms), potential barriers are formed, leading to a sudden decrease in mobility[28]. Thus, the decrease in the mobility values (Table SI.3) supports the hypothesis of the intercalation of Ga atoms.

Fig. 7a) shows a decrease of resistivity with decreasing temperature for the three studied samples, revealing a weak metallic behavior and suggesting that the transport is mainly dominated by bulk conductance. Undoped $Bi_2Se_3$ is more resistive than the Ga-doped samples for temperatures higher than 100 K. However, the decrease in resistivity is steeper towards 1.5 K for the former. The resistivity increases with doping percentage, which reveals a higher level



of disorder as the Ga concentration is increased. For undoped $Bi_2Se_3$, the electron concentration increases slightly with the temperature, contrary to Ga-doped $Bi_2Se_3$ samples, which show a slight decrease with increasing temperature (Fig. 7 b and c). These results are in agreement with the resistivity tendency. Linear behavior is expected for the resistivity, due to the electron-phonon scattering. The reduced slope for the doped samples is the consequence of a higher carrier concentration, which produces an increase in the screening of the deformation potential and therefore a reduced electron-phonon coupling. The electron mobility (Fig. 7 d) has a similar behavior for all doped samples. It decreases with temperature because the electron-phonon scattering will be more pronounced for lower temperatures, which reduces the drift velocity, reaching a maximum of around 1700 cm$^2$ cm$^2$V$^{-1}$s$^{-1}$ for undoped $Bi_2Se_3$. The variation in mobility with temperature for the doped samples is not so significant, because the carrier concentration deviation with temperature is not prominent.

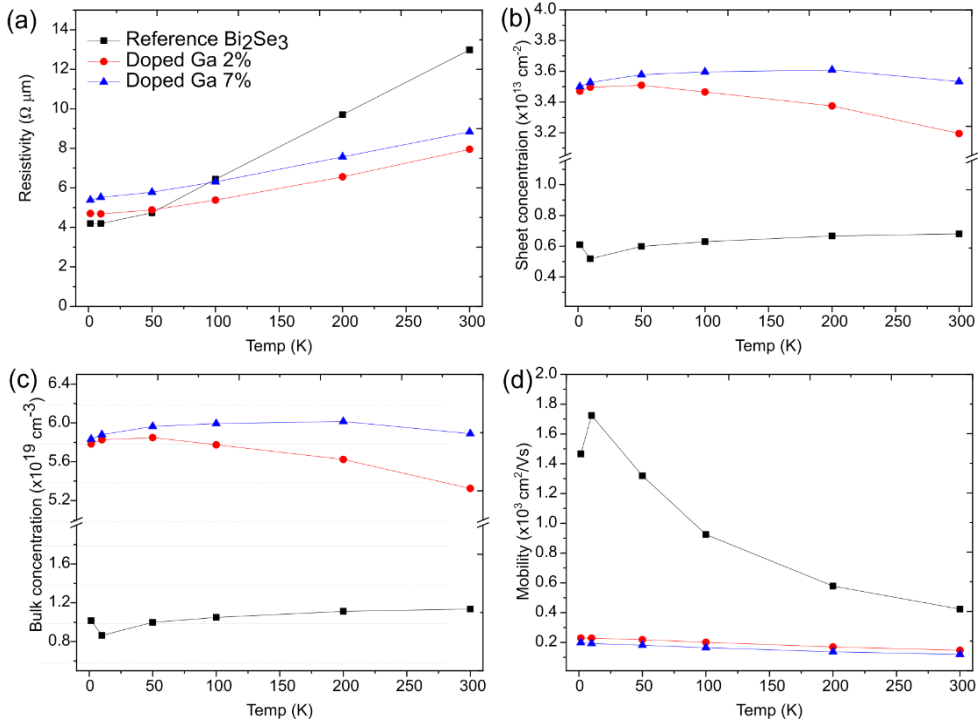

**Figure 7.** Hall effect measurements for undoped and doped $Bi_2Se_3$ with 2 and 7 at.% Ga doping. a) Resistivity, b) sheet concentration, c) bulk concentration and d) mobility as a function of the temperature. All samples are n-type.

At low temperatures, the transport properties enter a quantum regime, where the weak anti-localization (WAL) effect is manifested in TIs due to both the presence of strong-orbit coupling[29] in the bulk and 2D Dirac-like physics at the edges[30–32]. Spin-dependent scattering



due to either spin-orbit coupling in the bulk or spin-momentum locking at the surface leads to a quantum correction that enhances the conductivity. The magnetic field suppresses this correction, and the WAL arises as a sharp increase in the magnetoresistance with the magnetic field. The effect was observed in previous studies in $Bi_2Se_3$[8,33,34]. The presence of WAL in TIs can be described by the Hikami-Larkin-Nagaoka (HLN) model[29]:

$$\Delta G = G(B) - G(0) = \alpha \frac{e^2}{\pi h}\left[\ln\left(\frac{B_0}{B}\right) - \psi\left(\frac{1}{2} + \frac{B_0}{B}\right)\right] - \beta B^2, \qquad (2)$$

where $G$ is the magnetoconductance, $e$ and $h$ are the electron charge and Planck constant, respectively. B is the magnetic field, $B_0 = \frac{\hbar}{4eL_\phi^2}$, where $L_\phi$ is the coherence length and $\psi$ the digamma function. $\alpha$ is the prefactor, which is related to the number of conductive channels. The value of this parameter should be -0.5 per conductive channel because one topological surface state holds a $\pi$ Berry phase[35]. An ideal TI should present a prefactor of -1 due to the presence of two decoupled surfaces holding topological surface states. The $\beta B^2$ term accounts for the magnetic background[36], being $\beta$ the quadratic component. $\beta$, $L_\phi$ and the prefactor $\alpha$ are fitting parameters, which provide information about the quantum transport of the topological insulator[35].

Fig. 8 a) presents the magnetoconductance showing a symmetric WAL behavior for small magnetic fields. A low-pass filter was applied to remove high-frequency noise (S.I. 4). The WAL was studied to investigate the influence of the dopant on the quantum transport by fitting with the HLN model (dashed lines in Fig. 8 a). Fig. 8 b) shows the respective fitting parameters of the HLN equation as a function of the doping level. The sharp cusp in the magnetoconductance for the $Bi_2Se_3$ sample confirms the presence of conductive surface states. The fitting yields $\alpha \approx$ -0.96 , which is very close to -1. This value of $\alpha$ corresponds to two conductive channels in $Bi_2Se_3$ as seen in previous experiments [33,34,37–39,40], suggesting the presence of two decoupled surface states, the top, and bottom surfaces. The coherence length is ≈ 804 nm. For 2 at.% Ga, despite being less visible when compared to the reference sample, the cusp shows that the coherent transport of $Bi_2Se_3$ is maintained even at such doping level and with the reduction of the coherence length to 52 nm. This reduced value of $L_\phi$ is accompanied by a high value of $\alpha \approx$ -2.6, suggesting that the Ga doping leads to two extra channels with weaker phase coherence compared to the surface states[34]. S. de Castro et al.[41] observed a sudden transition of α from 0.7 to 2 at a critical temperature of 4 K in SnTe QWs, due to the creation of additional conduction channels. Shekhar et al. [42] proposed that a higher $\alpha$ can appear from a contribution of 3D conductance, due to the comparable dimensions of $L_\phi$ and the film thickness, electrons become aware of the 3D dimensionality of the film.

Similarly, WAL studies of different topological superconductor materials showed the presence of weak antilocalization, where higher values than -1 of the prefactor were observed for temperatures closer to the transition temperature[43,44]. Nevertheless, further theoretical, and experimental work is required to understand the electronic effect of intercalated Ga and to confirm if a superconducting phase can be observed.

On the other hand, 7 at.% Ga doping results in $\alpha \approx -0.07$ and $L_\phi = 717$ nm. The relevant reduction of $\alpha$ is likely due to the high concentration of Ga, which could lead to a disappearance of the surface states due to a topological transition, as it was observed for In doping at concentrations higher than 8 at.%[5]. The higher disorder produced by the high Ga concentration could also suppress the bulk contribution due to localization, thus explaining the reduced $\alpha$ value.

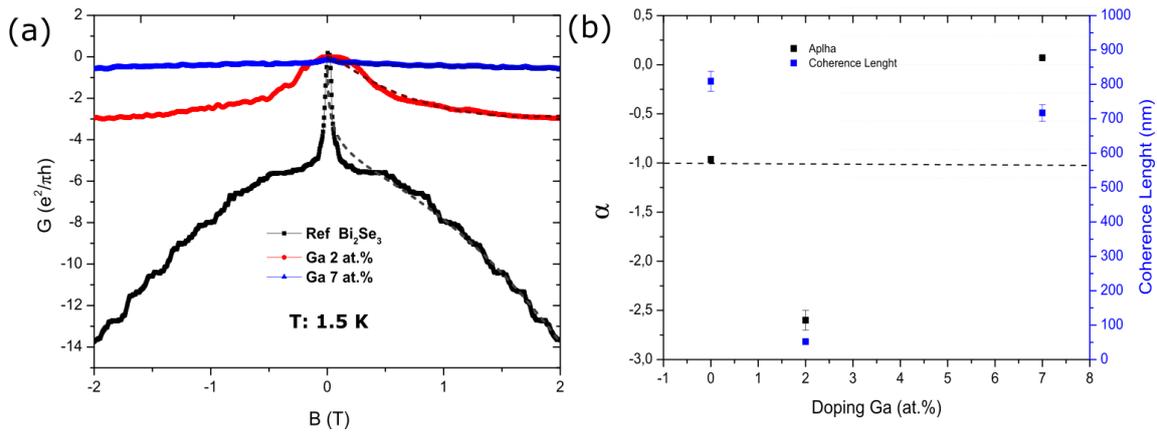

**Figure 8.** a) WAL measurements for undoped $Bi_2Se_3$ (black dots) and doped samples with 2 at.% (red dots) and 7 at.% (blue dots) Ga as a function of low magnetic field with the corresponding fittings by the HLN model (dashed lines). The sharp cusp is a typical feature of the presence of surface states. b) Fitting parameters $\alpha$ and phase coherence length ($L_\phi$) as a function of doping level.

## 4. Conclusions

The epitaxial growth of Ga-doped $Bi_2Se_3$ with various Ga concentrations is presented and the impact on the structural, vibrational, and electrical properties is discussed. XPS, Raman, and XRD measurements confirm the successful incorporation of the Ga dopants. The expansion of the crystal lattice and the redshift of the vibrational modes suggest that at low concentrations most Ga atoms intercalate in the van der Waals gap between the $Bi_2Se_3$ quintuple layers. The n-type doping nature provides additional evidence for the intercalation of Ga atoms.

As observed on intrinsic band insulator, the lack of WAL suggests the annihilation of the topological surface states as the Ga doping concentration is increased. On the other hand, for low doping, the atomic intercalation in the vdW layers and the increased WAL $\alpha$ are similar to previous reports for Cu, Sr, and Nb-doped $Bi_2Se_3$ [45–47], for which a speculated nematic superconductivity was observed. Therefore, our results for MBE-deposited Ga-doped $Bi_2Se_3$ open up the possibility of another material exhibiting topological superconductivity.

**Supplementary information**

See supplementary information for additional information.

**Acknowledgments**

This work was supported by Fundação para a Ciência e Tecnologia (FCT, Portugal) for the Ph.D. Grant SFRH/BD/150638/2020 and the INL internal grants SEED and EUREKA. We acknowledge the Agencia Estatal de Investigación of Spain Grant PID2019-106820RB and the Junta de Castilla y León Grant SA121P20 including funding by ERDF/FEDER. Carlos Tavares acknowledges funding from FCT / PIDDAC through the Strategic Funds project reference UIDB/04650/2020-2023 -

**Conflicts of interest**

The authors have no conflicts to disclose.

## Author contributions

**Daniel Brito:** Writing - Original Draft, Conceptualization, Methodology, Investigation **Ana-Pérez-Rodriguez**: Investigation, Writing - Review & Editing **Ishwor Khatri:** Investigation, Writing - Review & Editing **Mario Amado**: Investigation, Writing - Review & Editing **Carlos José Tavares:** Writing - Review & Editing **Eduardo Castro:** Writing - Review & Editing **Enrique Diez:** Writing - Review & Editing **Sascha Sadewasser:** Writing - Review & Editing **Marcel Claro:** Conceptualization, Writing - Review & Editing, Supervision

# Supplementary Information

**Effect of gallium doping on structural and transport properties of the topological insulator Bi$_2$Se$_3$ grown by molecular beam epitaxy**

Daniel Brito[1,2], Ana Pérez-Rodriguez[3], Ishwor Khatri[1], Carlos José Tavares[4], Mario Amado[3], Eduardo Castro[5], Enrique Diez[3], Sascha Sadewasser[1,6] and Marcel S Claro[1,6*]

7. International Iberian Nanotechnology Laboratory, Av. Mestre Jose Veiga, Braga 4715-330, Portugal
8. Departamento de Física e Astronomia, Faculdade de Ciências, Universidade do Porto, rua do Campo Alegre s/n, 4169-007 Porto, Portugal
9. Nanotechnology Group, Department of Fundamental Physics, University of Salamanca, 37008 Salamanca, Spain
10. Physics Centre of Minho and Porto Universities (CF-UM-PT), University of Minho, Azurém, 4804-533 Guimarães, Portugal
11. Centro de Física das Universidades do Minho e Porto (CF-UM-PT), Departamento de Físisca e Astronomia, Faculdade de Ciências, Universidade do Porto, 4169-007 Porto, Portugal
12. QuantaLab, 4715-330 Braga, Portugal

*corresponding author: marcel.claro@inl.int

## S.1 Ab initio lattice parameter

| Material | c-parameter (Å) | a-parameter (Å) |
|---|---|---|
| Pure Bi$_2$Se$_3$ | 28.115 | 4.1183 |
| Bi$_2$Se$_3$ (Ga 0.8 at.%) **Bi substitution** | 28.146 ↑ | 4.1100 ↓ |
| Bi$_2$Se$_3$ (Ga 0.8 at. %) **vdW intercalation** | 28.453 ↑↑ | 4.1306 ↑ |

Method: The *ab-initio* calculations were done using density perturbation theory (DFT) within the Perdew-Burke-Ernzerhof (PBEsol)[1] generalized-gradient-approximation (GGA) as implemented in Quantum Espresso[2,3] (QE) using Projector augmented wave (PAW) potentials. We have optimized lattice parameters and atomic positions using PBEsol in a 3×3×1 hexagonal supercell until forces in each atom are smaller than 0.05 eV/Å. The basis-set cut-off energy is 60 Ry and the Brillouin zone integrated with 3×3×3 Γ-centered Monkhorst-Pack grid of k-points, the self-consistent calculations with convergence criteria of 1×10$^{-8}$ eV.

## S.2 Quality of Bi$_2$Se$_3$

An atomic force microscopy (AFM) image for the reference Bi$_2$Se$_3$ sample is shown below. It is possible to observe a reduced twinning and large crystal size of almost 500 nm.

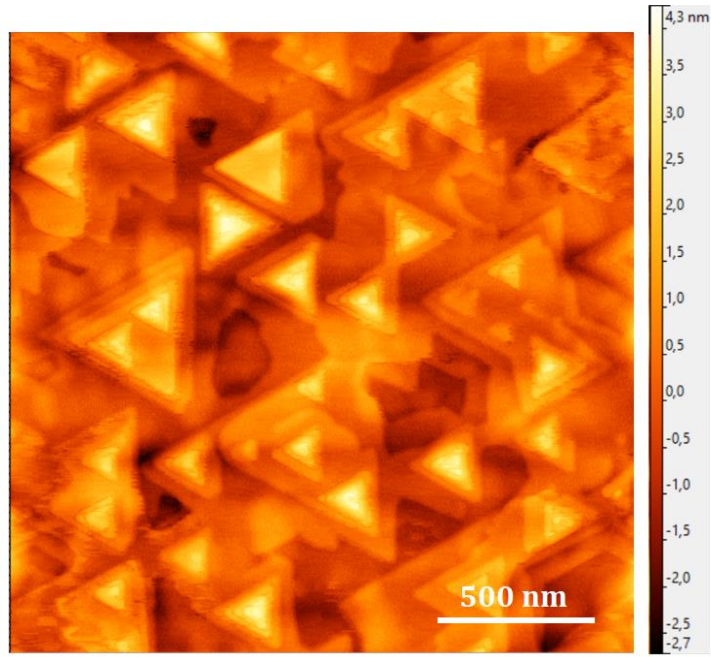

The diffraction peak (015) of Bi$_2$Se$_3$ was measured to perform the calculation of the lattice parameter a ($\mathring{A}$).

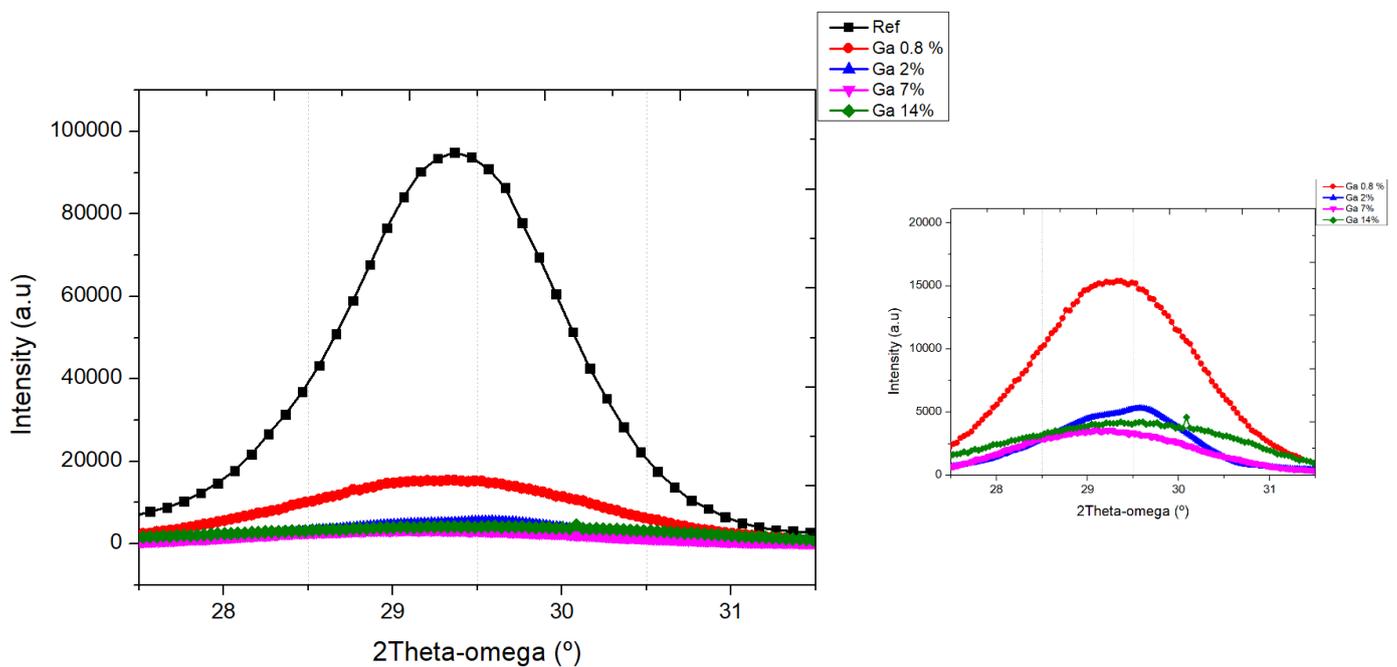

# S.3 Electrical properties at Room temperature

Table 1  Room temperature Hall measurements for all samples (n-type).

| Sample | Resistivity (Ohm μm) | Bulk Concentration ($\times 10^{19}$ cm$^{-3}$) | Sheet Concentration ($\times 10^{13}$ cm$^{-2}$) | Mobility (cm$^2$V$^{-1}$s$^{-1}$) |
|---|---|---|---|---|
| Bi$_2$Se$_3$ Reference | 12.98 | 1.13 | 6.81 | 423.53 |
| Bi$_2$Se$_3$:[Ga]=0.8 at.% | 10.91 | 2.82 | 11.27 | 203.03 |
| Bi$_2$Se$_3$:[Ga]=2 at.% | 7.95 | 5.32 | 31.94 | 147.42 |
| Bi$_2$Se$_3$:[Ga]=7 at.% | 8.84 | 5.89 | 35.33 | 119.94 |
| Bi$_2$Se$_3$:[Ga]=14 at.% | 18.50 | 3.27 | 19.66 | 103.2 |

# S.4 WAL filters and analysis

Application of lowpass filter to remove high frequency noise.

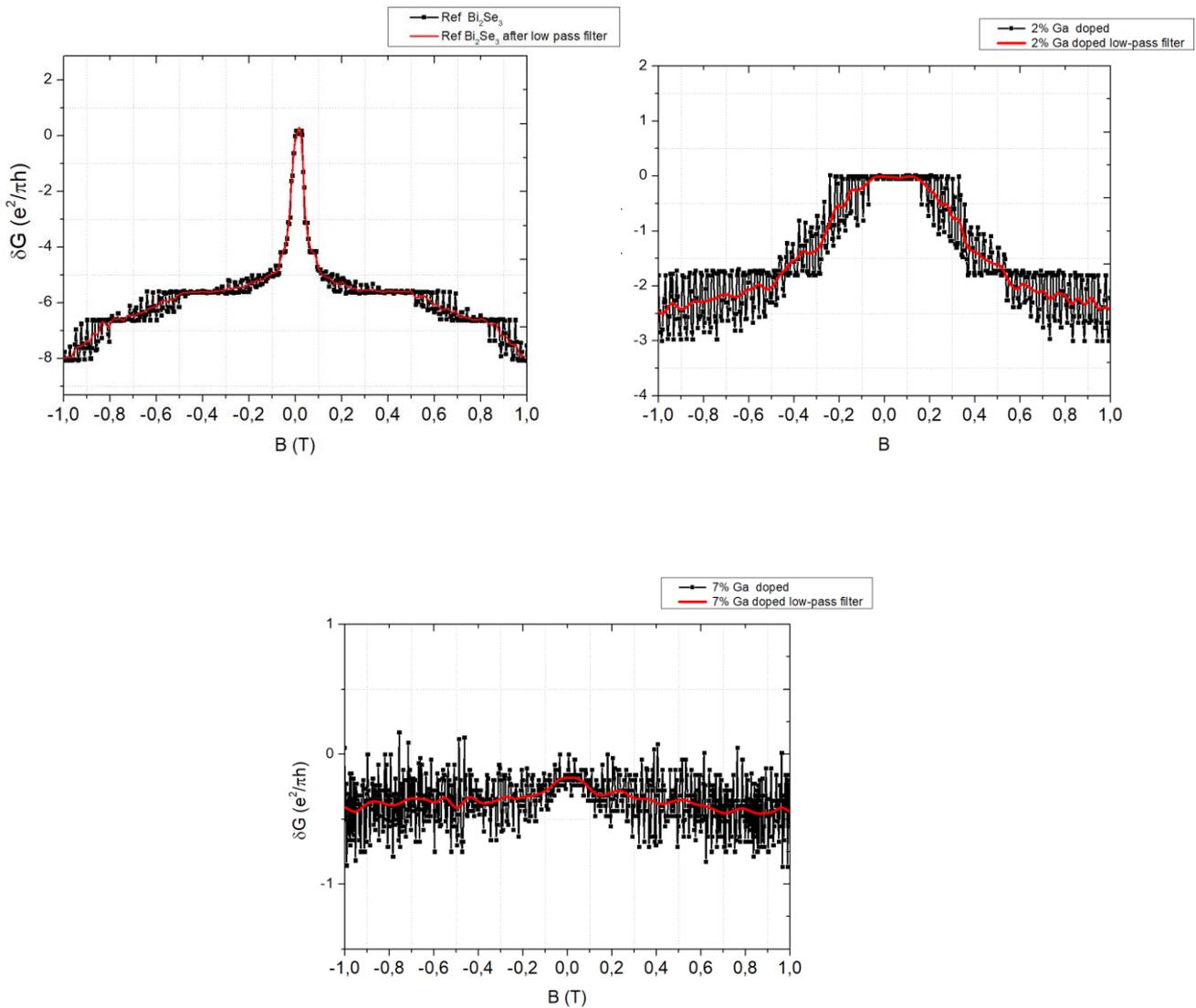